\documentclass[onecolumn,authoryear]{els-mrw} 
\usepackage{aas_macros,amsmath,amssymb,amsfonts,amsthm,makeidx,graphicx,txfonts,helvet}

\graphicspath{{./}{Figures/}} % look for figures in folder

\begin{document}

\chapter{Sagittarius A* - The Milky Way Supermassive Black Hole}\label{chap1}

\author[1]{Anna Ciurlo}%
\author[1]{Mark R. Morris}%
%\author[1,2]{Third Author}%

\address[1]{\orgname{University of California, Los Angeles}, \orgdiv{Department of Physics and Astronomy}, \orgaddress{Los Angeles, CA 90095, USA}}
%\address[2]{\orgname{Name of Institute}, \orgdiv{Division or Department}, \orgaddress{Address of Institute}}

\articletag{Chapter Article tagline: update of previous edition,, reprint..}

\maketitle

\begin{glossary}[Glossary]

\term{Active Galactic Nucleus (AGN)}
A highly energetic region at a galaxy’s center, powered by a supermassive black hole accreting matter.

\term{Advection-Dominated Accretion Flow (ADAF)} A type of radiatively inefficient accretion flow model in which most of the energy of the hot gas is advected (transported) into the black hole rather than radiated away.

\term{Accretion Flow} The process by which gas and other material spiral inward toward a black hole, forming a rotating disk. 

\term{Circum-Nuclear Disk (CND)} A dense ring of gas and dust surrounding Sgr~A* within a few parsecs, which acts as a reservoir for potential star formation and accretion onto the black hole.

\term{Eddington accretion rate}
The maximum rate at which matter can fall onto a compact object before radiation pressure from the accreting matter balances gravitational pull, limiting further accretion.

\term{Event Horizon}
The boundary surrounding a black hole beyond which the escape velocity exceeds the speed of light so that any matter or radiation is inevitably pulled toward the singularity at the center of the black hole. 
%The event horizon is not a physical surface but rather a mathematical boundary defined by the Schwarzschild radius for non-rotating black holes.

\term{Event Horizon Telescope (EHT)} A global array of radio telescopes that uses interferometry at high frequencies to achieve extremely high angular resolution. 

\term{Galactic Black Hole (GBH)}
The supermassive black hole at the center of the Milky Way Galaxy.

\term{Flare} A pronounced maximum in the continuously variable light curve emitted by accreting material in the region surrounding the event horizon of a black hole.
%sudden increase in brightness emitted by the region around a black hole, for Sgr~A* typically observed at infrared, radio, and X-ray wavelengths. 

\term{Frame Dragging} A relativistic effect in which the spin of a massive object, like a black hole, “drags” the surrounding spacetime with it. 

\term{Gravitational Redshift} A relativistic effect in which light emitted from within a strong gravitational field is stretched to longer, redder wavelengths as it escapes from the source of gravity. 

\term{Hot spot} A localized, heated region in the black hole’s accretion disk. They are thought to be heated through magnetic reconnection, emitting brief flares as they orbit near the Innermost Stable Circular Orbit (ISCO). %They provide insight into accretion dynamics and conditions near the event horizon.

\term{Innermost Stable Circular Orbit (ISCO)} The smallest orbit around a black hole where an object can stably orbit without spiraling inward. 

\term{Mini-spiral}
A triplet of ionized, orbiting gas streams within the central parsec of the Galaxy, consisting of bright, clumpy arcs that are projected on the plane of the sky to have the appearance of a 3-armed spiral.

\term{Power-law Distribution} A mathematical pattern observed in flares emitted by the region around a black hole, quantitatively describing how smaller flares occur more frequently than larger ones. 

\term{Quasi-Periodic Oscillations (QPO)} Short-term variations in light from an astronomical object that repeat briefly with near-constant frequency,  %, often observed in Sgr~A* flares and 
associated with hot spots or orbiting material near the innermost stable circular orbit (ISCO).

\term{Radiatively Inefficient Accretion Flow (RIAF)} A model describing low-radiation accretion flows, where most of the energy from infalling matter is carried inward through the event horizon before it can be emitted as electromagnetic radiation.

\term{Sagittarius A* (Sgr~A*)} 
The observational manifestation of the Milky Way galaxy's supermassive black hole (the GBH).
%name of the compact radio source arising from the immediate environment
%The designation for the supermassive black hole at the center of the Milky Way.

\term{Schwarzschild Precession} The gradual rotation of the major axis of the orbit of a star or object around a black hole, predicted by general relativity. 

\term{Schwarzschild radius}
The radius at which, an object of a given mass would become a black hole if it were compressed %if an object is compressed 
to a size smaller than or equal to it. %, the object would become a black hole. 
It corresponds to the radius of the event horizon of a black hole. 

\term{Supermassive Black Hole (SMBH)} A black hole with a mass ranging from millions to billions of times that of the Sun, typically found at the centers of galaxies.

\term{Synchrotron Emission} Radiation produced when charged particles moving at relativistic speeds, usually electrons, are accelerated by a magnetic field.
% spiral around magnetic fields at relativistic speeds. 

\term{Synchrotron Self-Compton (SSC)} A process in which synchrotron photons are scattered to higher energies by the same population of relativistic electrons that produced them. 

\term{Very Long Baseline Interferometry (VLBI)} A technique that combines signals from multiple radio telescopes around the world to achieve very high-resolution observations.

\term{Very Large Telescope (VLT)} Astronomical facility equipped with four 8 meter class telescopes located in Chile.

\term{Very Large Telescope Interferometer (VLTI)} A facility combining light from the telescopes of the VLT to achieve even higher resolution.

\term{Young Nuclear Star Cluster (YNC)}
A dense group of young (a few million years), massive stars (primarily O-type and Wolf-Rayet stars) orbiting the GBH within a radius of 0.5 pc. 

\end{glossary}

%\begin{glossary}[Nomenclature] %% COMBINED WITH GLOSSARY
%\begin{tabular}{@{}lp{34pc}@{}}
% EXAMPLE & etc etc etc \\
%\end{tabular}
%\end{glossary}

\begin{abstract}[Abstract]
This chapter provides a detailed overview of Sagittarius A* (Sgr~A*), the supermassive black hole at the center of the Milky Way, located in the dense Galactic Center region approximately 8 kpc from Earth. Despite its relatively low activity compared to more luminous active galactic nuclei, Sgr~A* has provided invaluable insights into black hole physics due to its proximity, enabling high-resolution observations of stellar orbits, gas dynamics, and variable emissions. 
In addition, Sgr~A* illustrates how supermassive black holes influence galaxy evolution through energy feedback and matter redistribution.
Early identification as a compact radio source and subsequent measurements of stellar orbits confirmed Sgr~A* as a black hole with a mass near 4 million solar masses. Observations of stars moving on tight, short-period orbits around the black hole have allowed direct tests of general relativity, such as gravitational redshift and orbital precession, under the influence of extreme gravitational fields. Sgr~A* displays variability across the electromagnetic spectrum, with flares in radio, infrared, and X-rays revealing complex interactions in the accretion flow, while outflows redistribute energy into the surrounding environment. Together, Sgr~A* and its environment offer a crucial window into the behavior of galactic nuclei. %, illustrating how supermassive black holes influence galaxy evolution through energy feedback and matter redistribution.
\end{abstract}

\paragraph{Keywords} Supermassive black holes --  Galactic Center -- Milky Way Galaxy -- General relativity -- Accretion -- Astrometry -- Radial velocity -- Light curves -- Massive stars

\paragraph{Learning Objectives}
\begin{itemize}
\item \textbf{Mass Measurement through Stellar Orbits} The orbits of stars around Sgr~A*, especially S0-2/S2, have provided an accurate measurement of the black hole mass, just over 4 million solar masses. This provided definitive evidence of the presence of a supermassive black hole at the Galactic Center, since nothing else could account for such a mass concentration.

\item \textbf{Testing General Relativity} Observations of S0-2/S2’s orbit near Sgr~A* allowed for critical tests of general relativity, including gravitational redshift and Schwarzschild precession. These tests have confirmed the theory’s predictions in an extreme gravitational field. %under extreme gravitational conditions.

\item \textbf{Multi-wavelength Variability} Sgr~A* shows significant variability across wavelengths—from radio to X-rays—with flares observed on short timescales. These emissions provide insights into the complex, turbulent processes in the magnetized accretion flow. % and reveal connections between different regions around the black hole.

\item \textbf{Impact on Galactic Environment} Although Sgr~A* is relatively inactive, it episodically produces outflows and extreme luminosity events that influence the surrounding circum-nuclear disk (CND) and potentially regulate star formation, illustrating the feedback role that even a quiescent black hole can have on its host galaxy. 
\end{itemize}

\section{Introduction: The Center of the Galaxy}
\label{sec:intro}
\begin{figure}[htb]
\centering
\includegraphics[width=\textwidth]{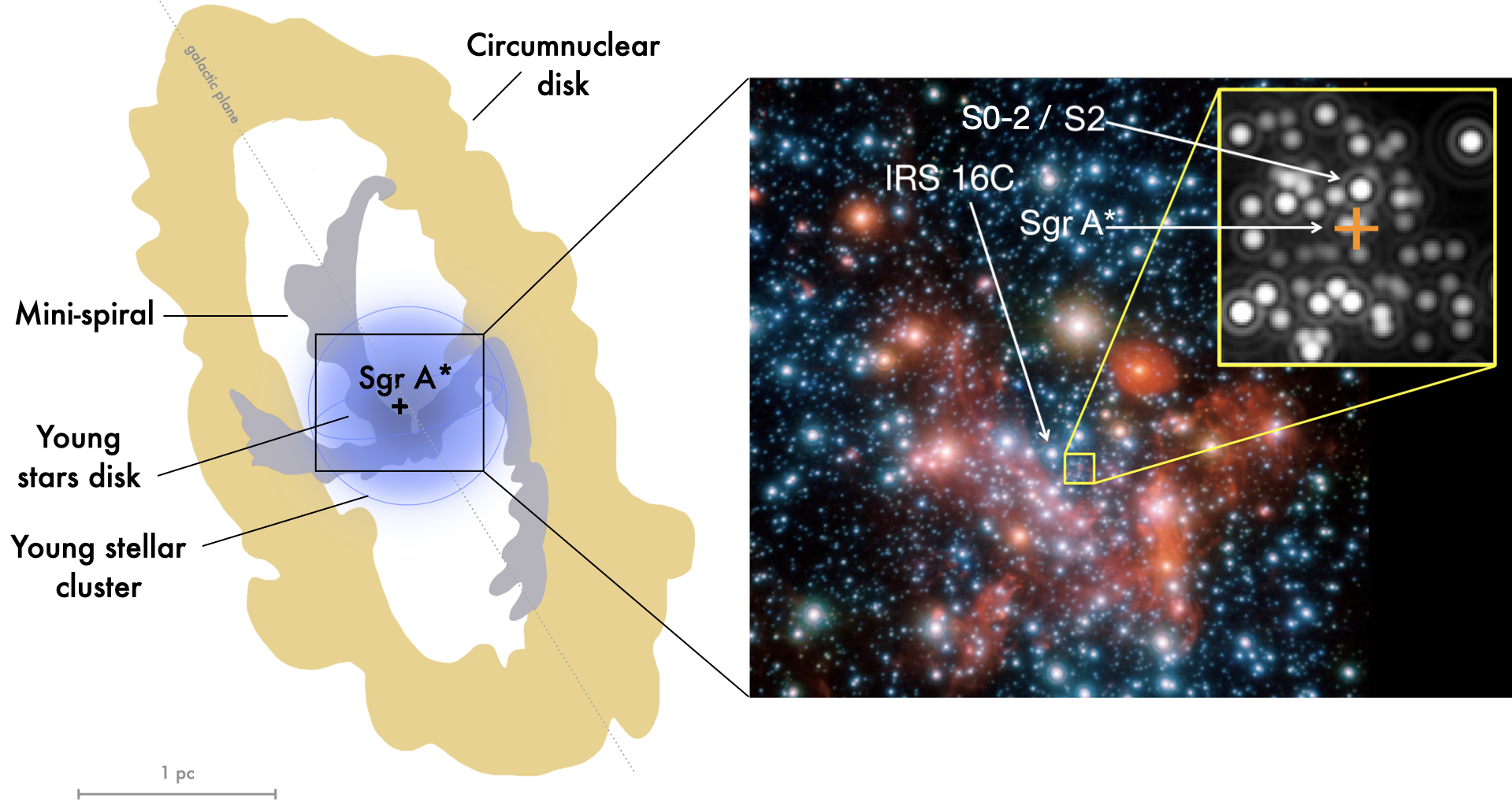}
\caption{Schematic of the central few parsecs of the Milky Way Galaxy. The two most prominent interstellar medium structures are the circumnuclear disk (in yellow) and the Mini-spiral (in gray). The radiation field is dominated by luminous young stars organized in the different sub-structures (shown in blue). The inset shows a zoom-in into the S-star cluster through an Infared image %orbiting tightly around Sgr~A* 
(inset credit: ESO/MPE/S. Gillessen et al.).}
\label{fig:GC-cartoon}
\end{figure}

The Galactic Center is a dense and complex region located about 8 kilo-parsecs from Earth, in the direction of the Sagittarius constellation. 
This region harbors a mix of old stars, young, massive stars, and dense molecular clouds, all orbiting a compact object, Sagittarius A* (Sgr~A*, see Figure~\ref{fig:GC-cartoon}). 
%Surrounding Sgr~A* is a cluster of stars known as the nuclear star cluster, one of the densest known star clusters.
The massive young stars are 4--6~Myrs old and many are organized into at least one coherently orbiting disk-like structure.
A separate, distinct population is the "S-star" Cluster, composed of a group of relatively young (6--20 Myrs) stars orbiting closely around Sgr~A* with high eccentricities and short periods.
A little further out (1.5 - 3 parsecs) Sgr~A* is also surrounded by the circumnuclear disk, a dense ring of gas and dust orbiting Sgr~A*, and within that, the Mini-spiral, consisting of three orbiting streams of ionized gas and dust.

\begin{figure}[htb]
\centering
\includegraphics[width=8cm]{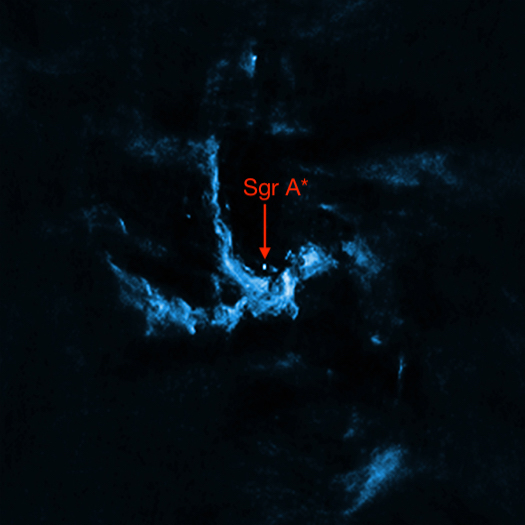}
\includegraphics[width=8cm]{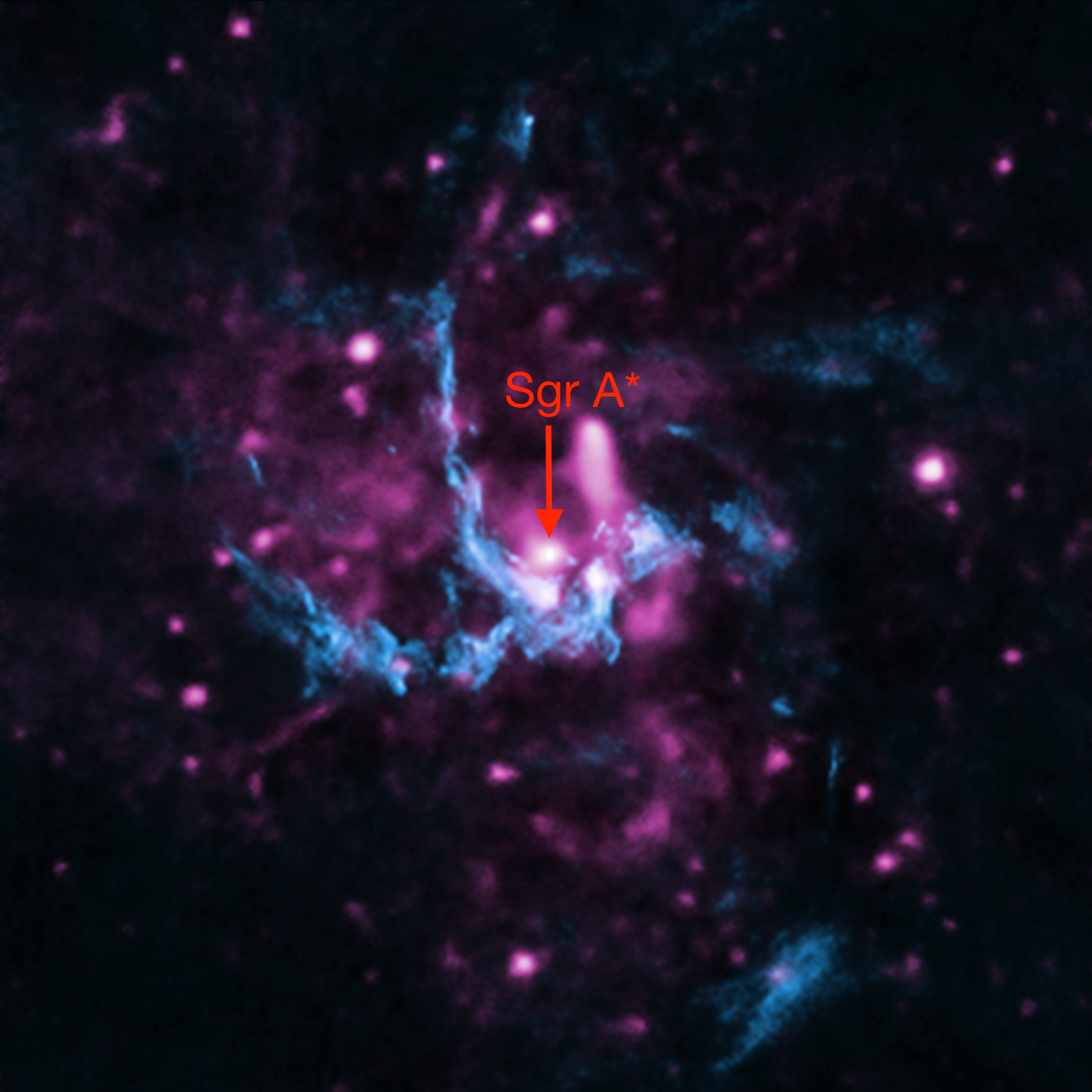}

\caption{The nucleus of the Milky Way Galaxy observed in the radio and X-rays. Left: image obtained with the JVLA radiotelescope at 1.3~cm with an angular resolution of 0.1~arcsec. Sgr~A* appears as the bright, point-like radio source in the middle of the image.  The extended emission surrounding Sgr~A* is the hot ionized gas within the central 16 arcsec region, often referred to as the Mini-spiral. 
This image was made from A and B array data at 1.3~cm \citep{1998ApJ...499L.163Z}. Right: X-ray map (shown in pink, \citealt{Li_2013}) superimposed on the same radio image. Sgr~A* shows up as an X-ray source. 
Both images are 1.2~arcmin across (about 3 parsecs).}
\label{fig:SgrA*_radio}
\end{figure}

The idea of a black hole at the center of the Milky Way began to take shape after advancements in infrared astronomy in the 1960s revealed a dense cluster of stars in the Galactic Center \citep{1968ApJ...151..145B}. 
Earlier radio observations had identified Sagittarius A (Sgr A) as a dominant radio-emitting complex \citep{1966ApJ...146..653D}, but it wasn’t until 1974 that \citet{1974ApJ...194..265B} used radio interferometry to discover a compact, bright source, later named Sagittarius A* (Sgr~A*), at the Galaxy's dynamical center (see Figure~\ref{fig:SgrA*_radio}). 
This source’s unique properties, including its brightness and rising energy distribution, suggested it was a candidate for a supermassive black hole, in line with predictions by \cite{1971MNRAS.152..461L}. In the late 1970s and 1980s, measurements of gas and stellar velocities hinted at the presence of a very massive object at roughly the location of Sgr~A*. However, the decisive demonstration that it is a black hole would require later evidence of a large mass confined to a small volume.

Despite being less active than some other supermassive black holes, Sgr~A* is the focus of a wide variety of research endeavors due to its proximity and therefore the opportunity it provides for detailed observations.

\section{The object at the Center}
\label{sec:GBH}
Early observations of Sgr~A* using Very Long Baseline Interferometry (VLBI) showed that the source was remarkably small: below 200~AU according to \cite{1975ApJ...202L..63L}, \cite{2005Natur.438...62S} pinned it down to about 1~AU. 
The large mass of the Galactic Black Hole (GBH) was first suggested by observations of gas motions around the Galactic Center, but it wasn’t until precise measurements of stellar orbits that astronomers could accurately estimate it. 
While early studies using gas clouds \citep{1979ApJ...227L..17L, 1985ApJ...293..445S} suggested a central mass of a few million solar masses, these estimates carried uncertainties due to non-gravitational forces affecting the gas. 
Stellar motions, on the other hand, provided a more reliable means of measuring the mass of Sgr~A*.

\begin{figure}[htb]
    \centering
    \includegraphics[width=7.4cm]{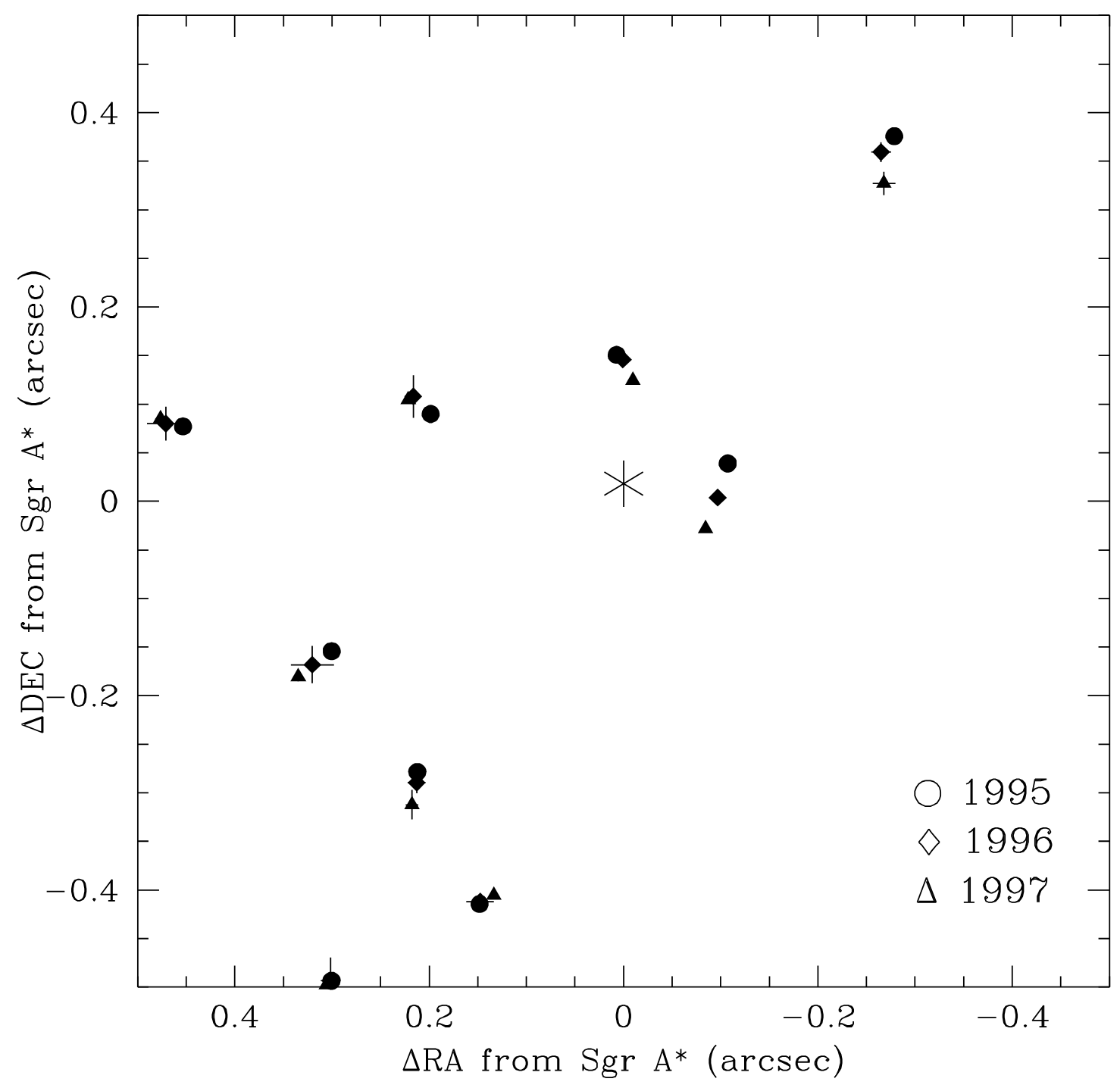}
    \includegraphics[width=3.35cm]{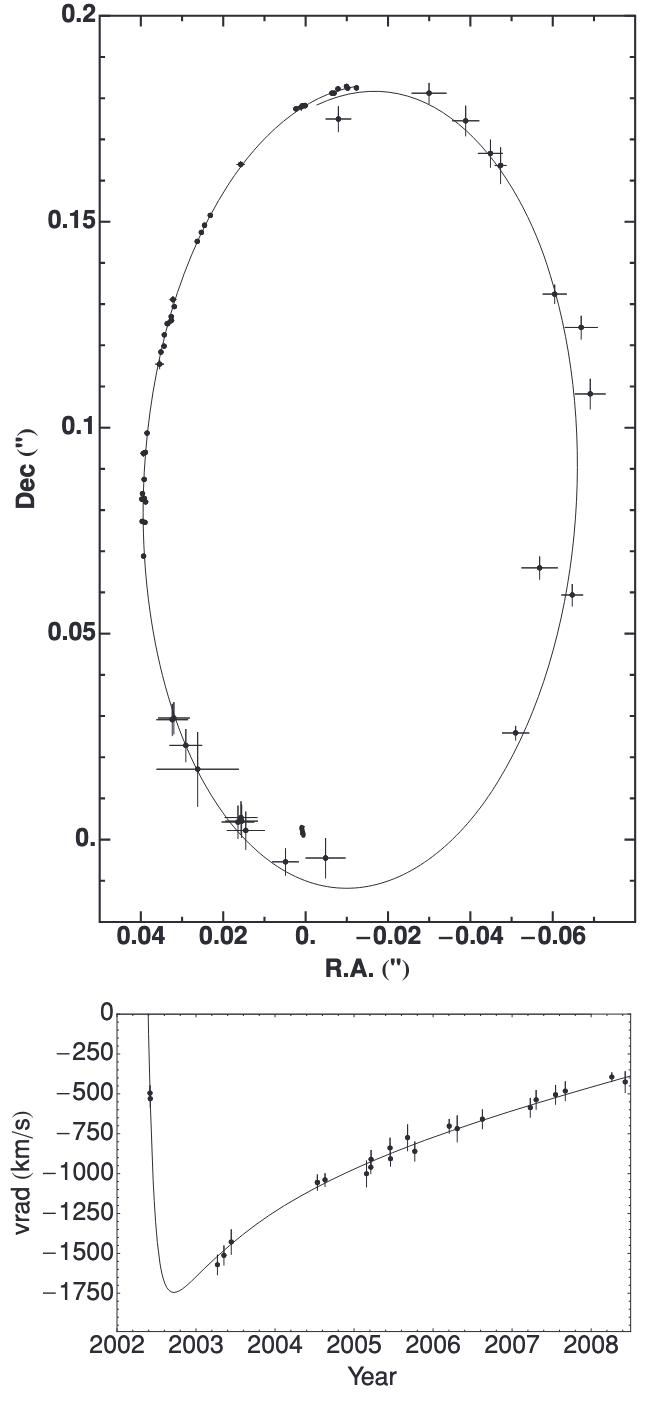}
    \includegraphics[width=5cm]{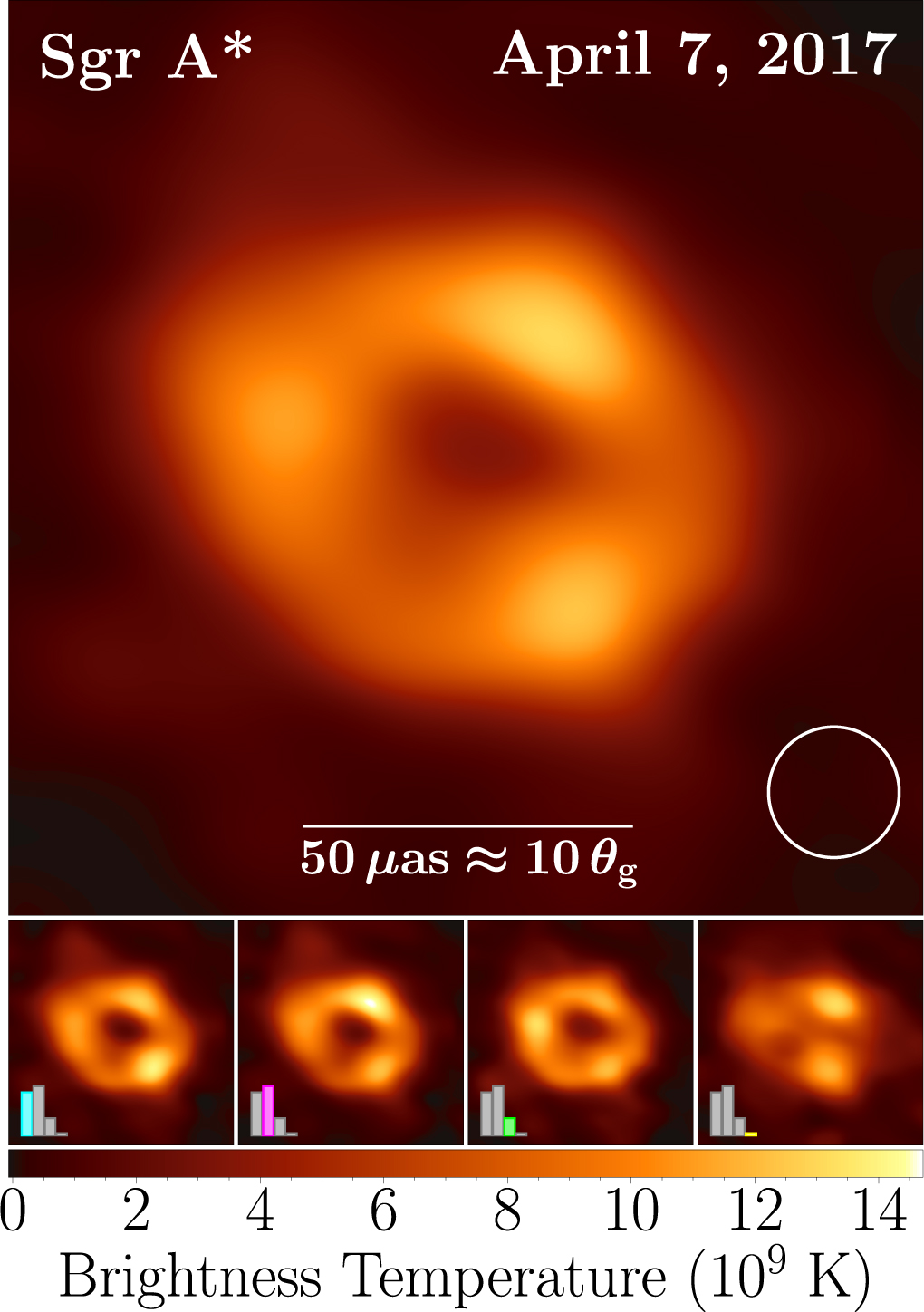}
    \caption{ Looking for the GBH.
    Left: measured proper motions of stars 1 arcsec$^2$ from the position of Sgr~A* (starred point, which depicts the location of the radio source). Credit: \cite{1998ApJ...509..678G}
    Center: S0-2/S2 measured astrometry (top) and radial velocities (bottom) fitted with a Keplerian model. Credit: \cite{2009ApJ...692.1075G}.
    Right: Representative EHT image of Sgr~A* from observations in 2017. This image is an average over different reconstruction methodologies. The bottom panels show average images within subsets having similar morphologies, with their frequency indicated by the inset bars. Credit: \cite{EHT_2022}.}
    \label{EHT}
\end{figure}

The breakthrough came when \cite{1997MNRAS.284..576E} and \cite{1998ApJ...509..678G} tracked the proper motions of stars near Sgr~A* -- their movement across the plane of the sky -- using high-resolution imaging techniques on 10-meter-class ground-based telescopes, the Very Large Telescope (VLT) and the W. M. Keck Observatory (Figure~\ref{EHT}, left panel). 
The proper motions of roughly 100 S stars  located within several arcseconds of Sgr~A* were measured, and the velocity dispersion as a function of radius was interpreted to imply the presence of a dark mass of at least 2.6 million solar masses confined within 0.1 arcsec of Sgr~A*.					
In addition to proper motion, acceleration measurements added another layer of precision. By tracking the change in the stars' velocities over time, astronomers were able to measure how strongly they were being pulled toward Sgr~A*. \cite{2000Natur.407..349G} reported the accelerations of three stars. 
%These acceleration measurements were key to ruling out alternative explanations, such as a dense but dark star cluster, and provided further evidence that a supermassive black hole.
Later, the time baseline of the astrometric measurements became sufficient to fit orbits \citep{2002Natur.419..694S, 2005ApJ...620..744G}. 
One star, S0-2 (also referred to as "S2"), was observed to complete an orbit around an unseen, massive object in just 16 years. 
%By fitting these stellar orbits to the observed data, astronomers estimated the mass of the object at around 4 million solar masses \citep{2008ApJ...689.1044G, 2009ApJ...692.1075G}.
Radial velocity measurements, which capture the stars’ motion toward or away from Earth using the Doppler shift of their spectral lines, further refined the mass estimate. 
This method added a third dimension to the stars’ motion, complementing the proper motion data. 
This allowed astronomers to construct full three-dimensional orbits for the stars (Figure~\ref{EHT}, mid  panel). 
For S0-2/S2, this allowed for an extremely accurate determination of the star's orbit and, in turn, the mass of the black hole: 4.1 million solar masses, with a high level of confidence (first measured by  \citealt{2008ApJ...689.1044G, 2009ApJ...692.1075G}, see \citealt{2018A&A...615L..15G, 2019Sci...365..664D, Kosmo_PhD} for the most up to date measurements).

Recently, the Event Horizon Telescope (EHT) released an "image" of Sgr~A* (\citealt{EHT_2022}, Figure~\ref{EHT}, right panel), revealing a bright ring surrounding the shadow of the GBH. This image is a reconstructed representation obtained by combining data from a global network of radio telescopes spanning multiple continents. This Very-Long-Baseline-Interferometry (VLBI) technique allowed the EHT to achieve the angular resolution necessary to observe features as small as the event horizon of Sgr~A*. 
The final image required complex modeling and computational algorithms to interpret the data. % and relies on assumptions about the accretion flow and black hole physics. 
The EHT team tried a variety of image reconstruction techniques, using a variety of assumptions about how to apply the techniques, and found general commonality in the resulting images. 
The resulting most likely image matched expectations for how the accretion flow should appear. 
This groundbreaking view of Sgr~A*’s immediate environment confirms its estimated mass, enabling direct tests of black hole physics at the Galactic Center.

Astronomers considered other exotic alternatives to a black hole. One possibility was a cluster of compact stellar remnants, such as neutron stars or white dwarfs, tightly packed into a small volume. 
However, such a cluster would be gravitationally unstable and would rapidly collapse into a black hole. 
Furthermore, the precise stellar orbits did not show any signs of interactions that would be expected if the mass were spread across a cluster rather than concentrated at a single point \citep{1998ApJ...494L.181M}.
Other hypotheses were a degenerate fermion ball, composed of massive particles like sterile neutrinos \citep{1998ApJ...500..591T}, or a boson star, a theoretical dense configuration of bosons \citep{2000PhRvD..62j4012T}. 
Both scenarios were ultimately ruled out due to the extreme density required to fit within the observationally constrained size of Sgr~A*. 
The compactness of the object, combined with the precise measurement of its mass, made these alternative explanations physically infeasible.

\section{Physics near the black hole}
\label{sec:GR}
\begin{figure}[htb]
    \centering
    \includegraphics[width=8cm]{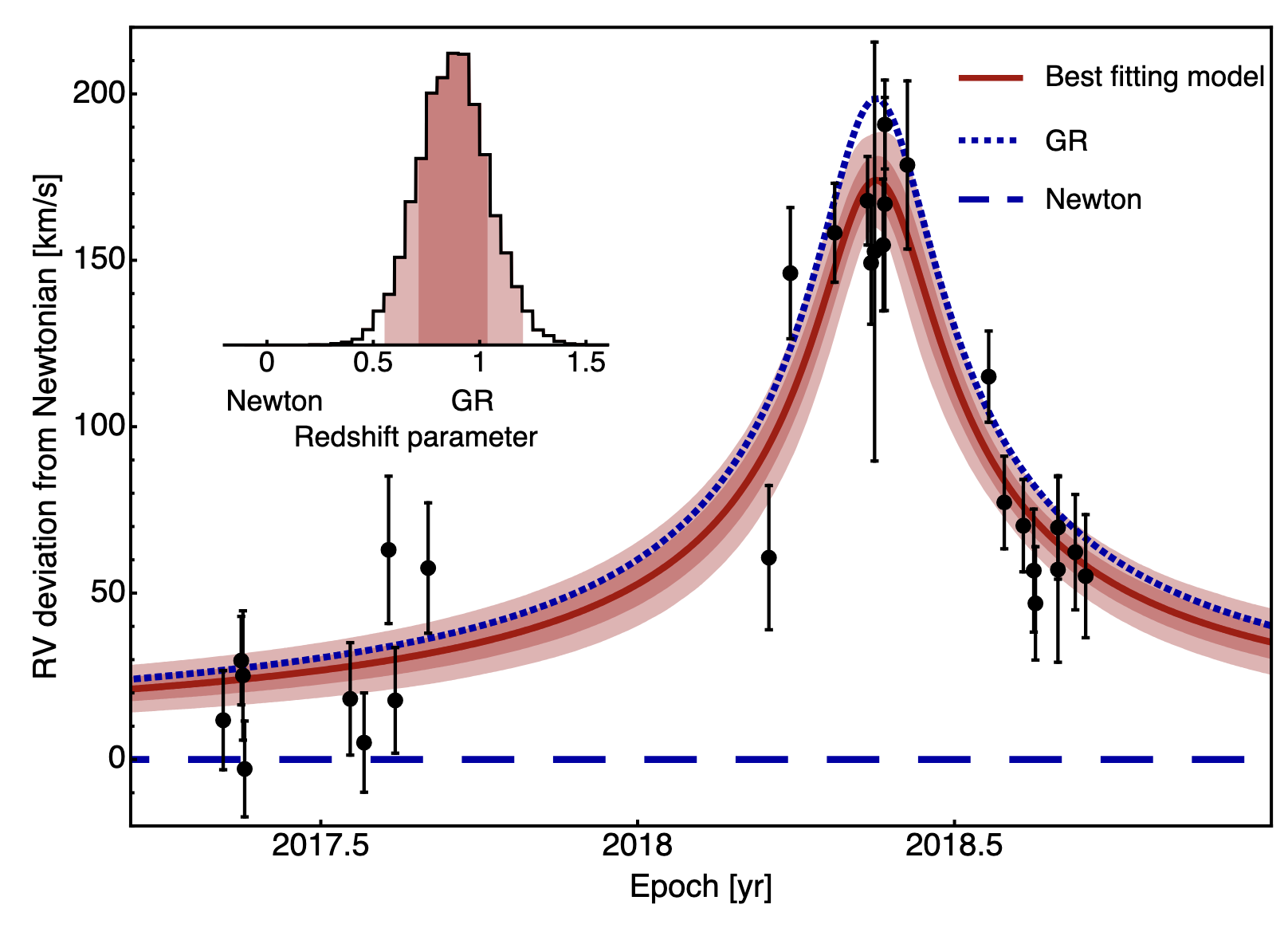}
    \includegraphics[width=8.2cm]{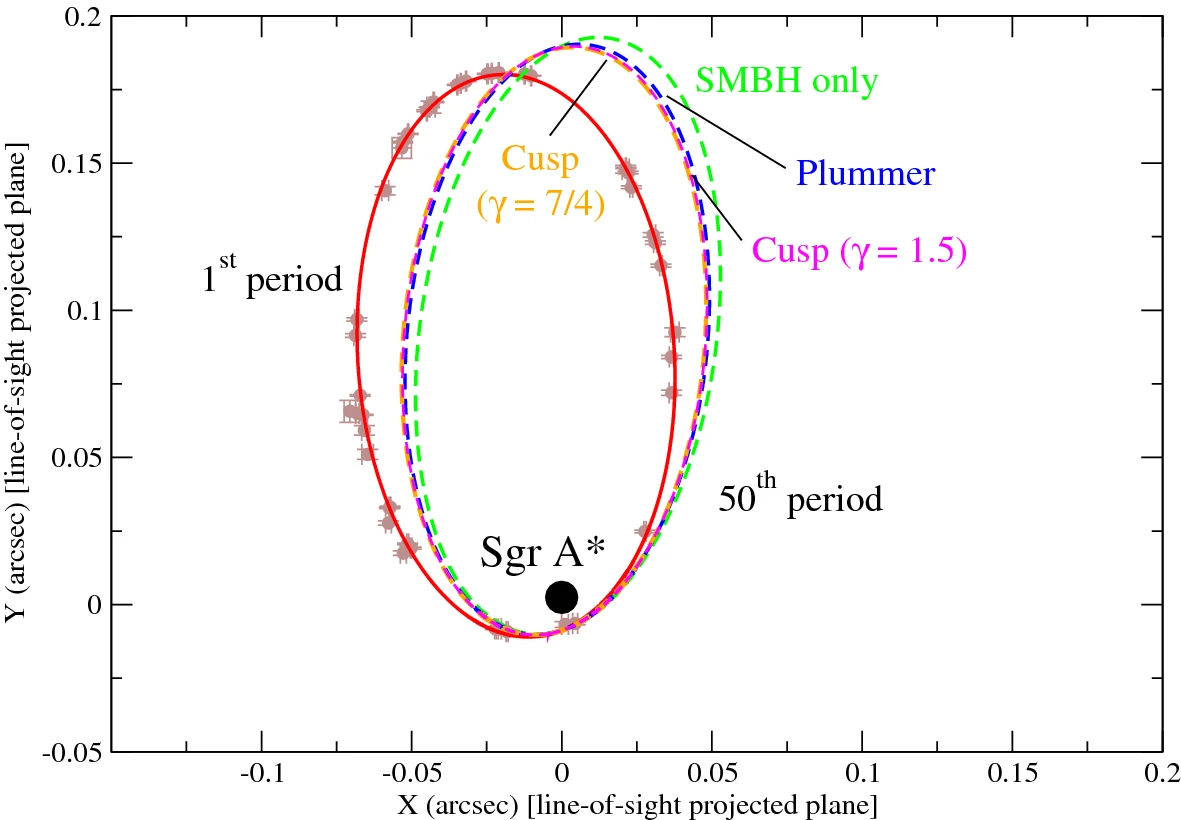}
    \caption{Testing physics near the GBH. 
    Left: Measured radial velocity deviation from Newtonian predictions. The best-fitting orbit model (red line) corresponds to $\Upsilon$=0.88$\pm$0.17. For comparison, the deviation expected for a purely relativistic signal is 1 (dotted line) and for a purely Newtonian model is 0 (dashed line). Credit: \cite{2019Sci...365..664D}.
    Right: Physics affecting the orbital  precession of S0-2/S2. 
    The best-fit orbit of S0-2/S2 data is shown in red. 
    The other ellipses show the predicted future orbits after 50 times the period, considering the presence of the SMBH alone (green) and the SMBH plus extended mass (magenta and blue, with different profiles).
    Credit: \cite{2022NatSR..1215258C}.}
    \label{GR}
\end{figure}

The GBH provides a unique opportunity to test general relativity in an extreme gravitational environment. 
One of the most critical tests came from observing the star S0-2/S2, which orbits Sgr~A* every 16 years. 
During its closest approach in 2018, S0-2/S2 reached speeds of about 2.7\% of the speed of light at a distance of 120~AU from Sgr~A*. 
This close passage allowed astronomers to measure its gravitational redshift, a relativistic effect in which light emitted by the star is stretched to longer wavelengths as it moves into the intense gravitational field near the black hole. 
To measure this redshift, astronomers tracked the star’s spectral lines, focusing on the combined Doppler and gravitational wavelength shifts as S0-2/S2 orbited closer to and farther from Sgr~A*. 
Both \cite{2018A&A...615L..15G} and \cite{2019Sci...365..664D} detected significant deviations from Newtonian predictions and confirmed the relativistic redshift effect with high precision. 
Figure~\ref{GR} shows the detection using the redshift parameter $\Upsilon$: its value is 0 in a purely Newtonian model and 1 in what is expected from general relativity.
This marked one of the most direct confirmations of general relativity in such an extreme environment.

Another important prediction of general relativity, more challenging to test, is orbital precession, whereby the star orbiting around a black hole does not return to the same position after completing one full orbit. 
Instead, the periapse (the point where the star is closest to the black hole) gradually shifts around the mass center, or precesses, over time, causing the star's orbit to slowly rotate around the black hole. 
In the case of S0-2/S2, two competing mechanisms affect this precession.
The first mechanism is Schwarzschild precession, which is a prograde (forward-moving) shift predicted by general relativity. 
This effect arises from the curvature of spacetime near the black hole. 
%Observations from the \cite{2020A&A...636L...5G} confirmed that S0-2/S2's orbit exhibits this relativistic precession, matching general relativistic predictions.
This effect is made even more challenging to measure by the second mechanism, Newtonian precession, which can be caused by the presence of extended mass, such as dark matter, stars, or stellar remnants surrounding the supermassive black hole. 
This effect would induce a retrograde (backward-moving) precession, counteracting the Schwarzschild precession (see Figure~\ref{GR}. 
The strength of this effect depends on the amount of extended mass within the minimum and maximum radii of the star's orbit. 
The initial investigations of this effect to date have reported that the measured precession is consistent with pure general relativity, indicating that any extended mass within the orbit of S0-2/S2 must be far less %at least 1000 times less 
than the mass of the black hole itself \citep{2020A&A...636L...5G}. 
This places strong constraints on the amount of unseen matter in the immediate vicinity of the black hole (up to 1200 solar masses within the orbit of S0-2/S2; \cite{2024A&A...692A.242G}) which is about 3000 times less than the black hole mass.

Future observations of stars with tighter orbits around Sgr~A* may allow astronomers to refine this measurement and possibly detect even more subtle relativistic effects, such as frame dragging, where the black hole’s spin causes spacetime itself to twist around it. 
Although these effects are presently much more difficult to observe, improvements in technology, along with continued monitoring, will allow for the exploration of such effects by monitoring the stars orbiting more tightly around Sgr~A* so that may soon provide further insights into the extreme physics near the Galactic Center.
If a pulsar could be found orbiting closely around the GBH, its potential for precise timing experiments could be the most promising way to investigate general relativistic predictions around a black hole \citep{Pfahl+Loeb04, Torne+23} 

\section{Star formation near the black hole}
\label{sec:starformation}

The presence of a cluster of $\sim$100 massive, young stars $\sim$0.5~pc around the GBH -- the Young Nuclear Cluster (YNC, \citealt{LuJR18}) -- originally led to skepticism that a massive black hole could be present at that location because, as was argued by \citet{Sanders92}, the tidal force of a SMBH %black hole having the mass inferred at that time by dynamical measurements 
would have inhibited the formation of stars.  
The reason is that, in order for a potential protostellar cloud to exceed the Roche density -- needed for the cloud to be self-gravitating against the disrupting tidal forces -- the density would have to have been at least 5 orders of magnitude larger than anything that has been measured in the central parsecs of the Galaxy.  
%When the black hole was later demonstrated to be present, 
This situation was termed the "Paradox of Youth".  

\subsection{Impact of star formation near Sgr~A*}
The age of YNC has been constrained to be $5\pm2\times10^6$ years by the presence of a few dozen Wolf-Rayet stars and OB supergiants \citep{Krabbe+95, LuJR+13}.  
The gas out of which the YNC formed has almost entirely dissipated; only a few tens of solar masses of gas is left in the volume occupied by the cluster.  
%Anna: this is true for the ionized gas, but is this statement still true for molecular hydrogen?
The formation of the YNC must therefore have occurred under conditions far different from those we find there now.  
First, the cloud or disk out of which the YNC formed must have had a mass of several tens of thousands of solar masses, with particle densities ranging up to $10^{11}$ cm$^{-3}$, 4 or 5 orders of magnitude larger than anything observed in the region at present. 
Second, the formation of this $\sim10^4$ solar mass cluster in the immediate vicinity of the black hole would create outflows -- jets, stellar winds, supernovae -- that would push material in all directions, including onto the SMBH.
Therefore, this star formation event could not have occurred without causing accretion onto the GBH at a far higher rate than we see now, possibly approaching the limiting Eddington accretion rate.
At such a high rate of accretion, the Galactic Center would have been an active galactic nucleus (AGN) %, likely having the characteristics of a Seyfert galaxy nucleus, including 
with a high luminosity in ultraviolet radiation. 

Evidence for such a past AGN phase, compatible with the formation time of the YNC, has been found in the Magellanic Stream of atomic hydrogen, which stretches over the southern pole of the Galaxy in the orbital wake of the Magellanic Clouds.
\citet{Bland-Hawthorn+13} pointed out, using observations of H$\alpha$ line emission, that the Stream has been ionized in the section that would have been exposed to the strong ultraviolet radiation emanating from the then presumably active Galactic Center.  
They %also 
estimated that the time needed for the ionized hydrogen %in the Stream 
to recombine and relax to its currently observed H$\alpha$ surface brightness 
is %on the order of a few million years, 
consistent with the age of the YNC.
Another line of evidence for an %Seyfert
outburst from the Galactic Center several million years ago is the presence of the Fermi Bubbles (see Section~\ref{subsec:outflow}) which, in one interpretation, were produced by strong, collimated outflows carrying relativistic particles from the GBH during a high-accretion AGN stage several million years ago \citep{Zubovas+11, Zubovas+12b, Yang+22}, consistent with the age of the YNC.

\subsection{Origin of the young stars near Sgr~A*}
The question remains how star formation could have taken place in the central light-year of the Galaxy.  
Many of the stars in the YNC are distributed in a disk-like configuration \citep{LevinBeloborodov03, Paumard+06, Yelda+14, 2023ApJ...949...18J, 2022ApJ...932L...6V}, 
suggesting that there must have been a massive disk of gas orbiting closely around the GBH.  
Such a circumnuclear disk (the CND) exists at the present time, but it has an inner cavity of $\sim$1.5 parsec radius, substantially further out from the black hole than the extent of the YNC, so in this scenario the disk must previously have extended much closer to the GBH, as mentioned above.  
A simulation of how stars might have formed in such a massive, high-density disk was developed by \citet{Nayakshin+07}, using various simplifying assumptions.  Those authors found that %the initial mass function (IMF) 
in this extreme situation, % would favor 
massive stars would be favored 
compared to %the standard IMF that generally describes 
star formation elsewhere, %a result that
which is %is in 
qualitatively in agreement with observations of the YNC \citep{LuJR+13}.  The computational problem of modeling the formation of the YNC will be very complex when all of the physics is eventually included (magnetic fields, feedback from both the stars and the black hole, a complete treatment of heating and cooling).  
This will be needed, however, to convincingly solve the Paradox of Youth.  

Another class of models for star formation around the GBH is based on an assumed collision between the GBH and a massive cloud that is falling toward it. There are numerous published investigations of this basic idea \citep[][and many others listed by \citet{Dinh+21}]{Sanders98, Bonnell+Rice08, Generozov+22}.  This scenario leads to the formation of a dense disk in which  stars can subsequently form. However, \citet{Dinh+21} argue that this scenario is unlikely because it is very difficult to arrange for clouds to fall toward the bottom of the Galaxy's gravitational potential well on near-zero angular momentum orbits.

The CND is an obvious potential reservoir for future star formation around the GBH.  As the CND undergoes viscous evolution, promoted by its strong internal turbulence and strong magnetic field, its inner edge will migrate inward over time until it reaches the GBH, and the next round of star formation and AGN activity might be expected to ensue.  As long as the CND is quasi-continuously replenished by inward migration of material from the Central Molecular Zone \citep{Tress+20}, this could be a cyclical phenomenon \citep{1999AdSpR..23..959M}, not just in the Milky Way, but in gas-rich spiral galaxies in general.   

Star formation takes place continuously on larger scales in the Galactic Center, where the tidal force from the GBH is much reduced, as reviewed by \citet{Henshaw+23} and by \citet{Morris23}.  
However, whether star formation is presently ongoing in the tidally affected central parsec around the GBH is currently under debate \citep{Morris23}.

\section{Black hole activity}
\label{sec:activity}
The GBH is known for its low luminosity, but it exhibits pronounced variability across the electromagnetic spectrum, from radio to X-rays. These fluctuations occur on timescales ranging from minutes to hours and provide valuable insight into the physical processes driving the accretion flow and emission mechanisms around the black hole. 
%By observing Sgr~A* across multiple wavelengths, astronomers have been able to build a more complete picture of its behavior and constrain theoretical models of accretion.

\subsection{Short term variability}
\label{subsec:shortterm}
\subsubsection{Observed Variability Properties}
\label{subsubsec:obs}
\begin{figure}[htb]
    \centering
    \includegraphics[width=\textwidth]{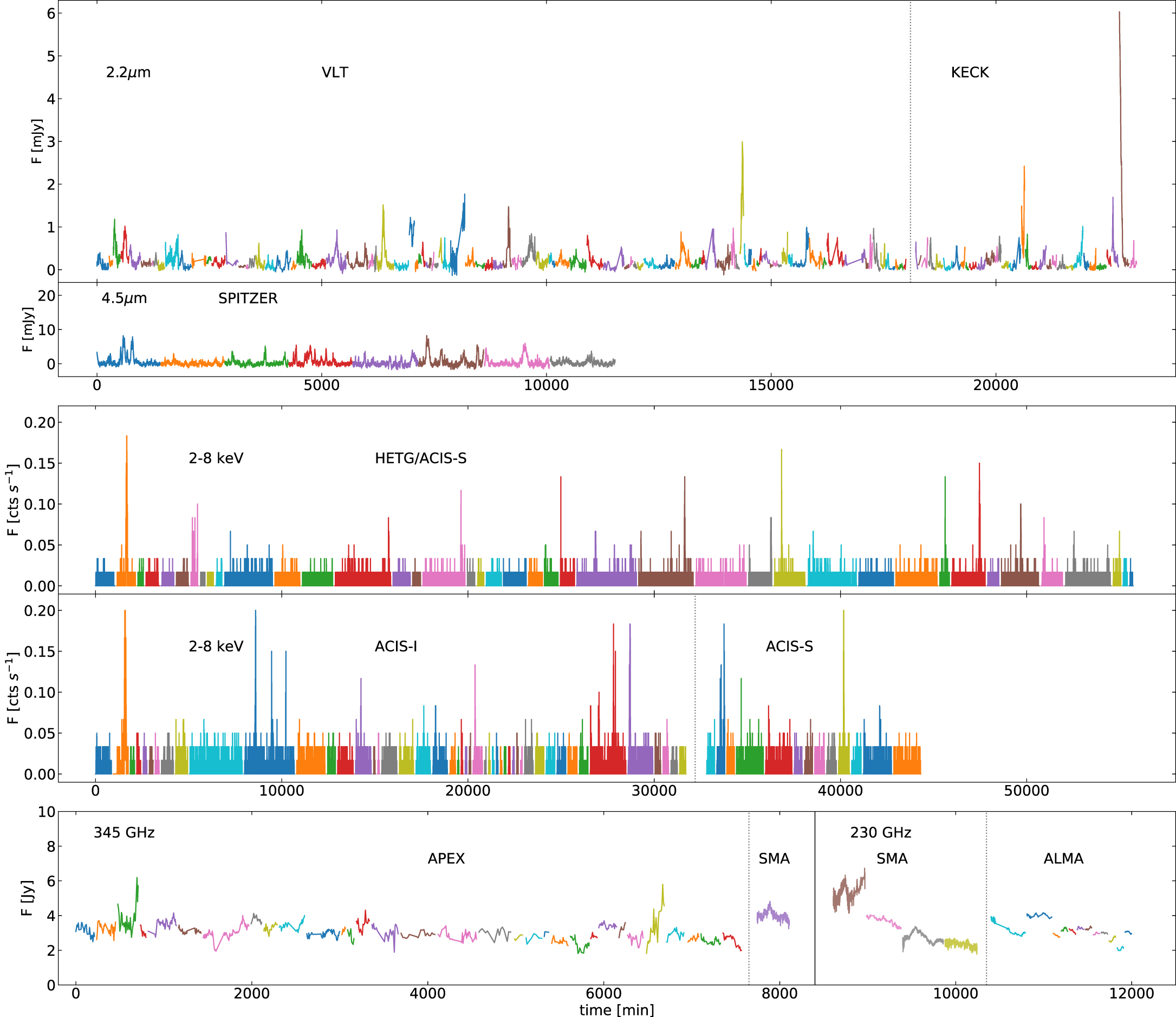}
    \caption{Observed light curves for Sgr~A* at different wavelengths. The data are presented without observational gaps between the epochs (hours to years), and each epoch is shown in a different color (same color does not indicate common epochs across panels). 
    From top to bottom: infared data (2.2~$\mu$m) from the VLT and Keck,  
    mid-infrared data (4.5~$\mu$m) from Spitzer, X-ray data from three different detectors on the Chandra X-ray Observatory, and finally radio data obtained with APEX/LABOCA, SMA and ALMA (345~GHz and 230~GHz).
    Credit: \cite{2021ApJ...917...73W}.}
    \label{variability}
\end{figure}

The variability of Sgr~A* has been observed across a broad range of wavelengths  (Figure~\ref{variability}), each probing different regions of the accretion flow.
Radio wavelengths trace the extended, lower-energy regions of the accretion flow. 
Observatories like the Jansky Very Large Array (JVLA), the Sub-Millimeter Array (SMA), and the Atacama Large Millimeter Array (ALMA) monitor Sgr~A* from centimeter to sub-millimeter wavelengths, where variability is detected over hours to days.
Infrared observations, primarily with the VLT and Keck Observatory, provide high-resolution data on synchrotron emission from hot electrons in the inner accretion flow, often associated with rapid flares.
X-ray observations from Chandra and XMM-Newton capture the highest-energy emissions, offering a glimpse into the most energetic processes occurring near the black hole. 

Sgr~A* shows continuous, random variability across a broad range of timescales. 
The emission process is highly dynamic, with fluctuations happening continuously, including flares -- prominent peaks in brightness that occur several times a day. 
The distribution of flare amplitudes follows a power-law pattern characterized as "red noise" on time scales up to several hours \citep{Mauerhan+05,2009ApJ...691.1021D, 2018ApJ...863...15W}, meaning that small-amplitude fluctuations occur more frequently than large ones, but without a characteristic or dominant timescale. 
This suggests that the underlying processes driving the variability are scale-invariant, possibly related to turbulence, shocks, or magnetic reconnection within the accretion flow \citep{2018ApJ...863...15W}. 
The variability could also reflect variations in the instantaneous accretion rate resulting from dynamical instabilities in the accretion flow.
This power-law behavior has been observed across the infrared, sub-millimeter, and X-ray regimes, providing important constraints on models of accretion and emission near the black hole.

In the radio and sub-millimeter, variability is moderate, with flux-density changes on the order of 20-30\% over timescales of hours to days \citep{YZ+06,2021ApJ...917...73W}. 
In the infrared, Sgr~A* exhibits more rapid and pronounced variability, with bright flares having rise times of minutes and typical durations of 20-30 minutes (see Figure~\ref{variability}. %occurring on timescales of minutes. 
In the X-ray regime, Sgr~A* displays detectable flares about once per day, some of which can be rather intense, with flux increases over the quiescent background by factors of up to 600 \citep[e.g.,][]{Haggard+19}. 

One of the key challenges in understanding Sgr~A*’s variability is determining how flares across different wavelength regimes are related. 
X-ray flares are almost always observed to occur nearly simultaneously with near-infrared flares \citep{BoyceH+19}. 
However, many infrared flares are not accompanied by a simultaneous X-ray flare.
At radio and sub-millimeter wavelengths, the correlation with higher-energy flares in the infrared and X-rays is less straightforward. 
Some sub-millimeter maxima have been observed to follow infrared peaks with a delay of 0.5 - 3 hours \citep{2008ApJ...682..373M, 2012RAA....12..995M, Michail+21, BoyceH+22}, possibly due to the time it takes for material to propagate outward from the inner accretion flow. 
However, the time delay is not well established in a statistical sense, and it remains possible that at least some of the delayed sub-mm maxima that have been reported are coincidental and unrelated. 
The complex, multi-wavelength behavior of Sgr~A*’s variable emission implies that different physical processes dominate at different wavelengths, as discussed in the next section. %~\ref{models}, 
%and that the flares may represent localized events such as magnetic reconnection or shocks.

\subsubsection{Theoretical Models and Interpretation}
\label{models}
The observed variability of Sgr~A* is generally explained by radiatively inefficient accretion flow (RIAF) models \citep{1995ApJ...452..710N,2014ARA&A..52..529Y}. 
In these models, the low luminosity of Sgr~A* is due to the inefficient conversion of gravitational energy into radiation, while flares result from localized, transient events within the accretion flow.

Infrared  flares are believed to consist of synchrotron emission from relativistic electrons in the inner accretion disk or from orbiting "hot spots" near the black hole’s innermost stable circular orbit (ISCO).
As material approaches the event horizon, its energy increases dramatically, causing many particles to move at relativistic speeds (close to the speed of light). These high-energy particles interact with the magnetic fields generated by the accretion flow, producing synchrotron emission. Magnetic reconnection events can trigger sudden energy releases, creating localized hot spots in the accreting material \citep{Dexter+20, Ripperda+20, Ripperda+22}. Similarly, the colliding winds from the massive stars in the YNC can also form gas clumps within the accretion flow. When these clumps are heated to extreme temperatures, they transform into hot spots where particles become relativistic and emit synchrotron radiation.
Inside the ISCO, any orbit becomes unstable, meaning that any material or object would quickly spiral inward toward the black hole. The location of the ISCO depends on the black hole's mass and spin: for a non-rotating black hole, it occurs at 3 times the Schwarzschild radius, but for rotating black holes, the it can be closer to the event horizon. The ISCO is critical for understanding the dynamics of accretion disks, as material within this orbit rapidly plunges toward and through the event horizon \citep{2003Natur.425..934G, 2019ApJ...882L..27D}.
Hot spot models propose that quasi-periodic infrared flares result from clumps of plasma orbiting near the ISCO \citep{2004ApJ...606..894Y}.
Observations reported by \cite{2020A&A...635A.143G} revealed quasi-periodic oscillations (QPOs) with periods of around 20-30 minutes, which are consistent with material orbiting near the ISCO \citep{2022A&A...665L...6W}\footnote{On short time scales hot spots can repeat at the orbital period and only last a few orbits.}.

The temporal correlation between infrared and X-ray flares suggests that the emission mechanisms in these two frequency regimes are closely linked and arise from the same population of relativistic electrons.
X-ray flares are most often explained as resulting from synchrotron self-Compton (SSC) emission, whereby synchrotron photons produced in the infrared by the relativistic electrons are Compton scattered to X-ray energies by 
%higher-energy 
the same population of electrons \citep{2001Natur.413...45B,2006ApJ...650..189Y,2013ApJ...774...42N}. 
However, not all infrared flares are accompanied by X-ray flares, indicating that the electron energy distribution produced in a flare does not always extend to sufficiently high energies to scatter photons up to X-ray energies \citep{2021ApJ...917...73W}.
%only a fraction of the synchrotron emission produces the conditions necessary for SSC scattering.
%The rapid rise and fall of X-ray flares indicates that they occur in the innermost regions of the accretion flow, within a few Schwarzschild radii of the black hole.
The X-ray flare's rapid rise and fall (minutes to hours, \citealt{2001Natur.413...45B}) aligns with the orbital periods of matter near the innermost regions of the accretion flow, within a few Schwarzschild radii of the black hole \citep{2023A&A...669L..17V}. 
%The intense magnetic fields and extreme conditions in this region make it ideal for processes like magnetic reconnection, hot spot formation, and particle acceleration, which are thought to produce the flares. 

Radio variability is often less pronounced and occurs on longer timescales, suggesting that it originates from more extended regions of the accretion flow. 
Sub-millimeter variability appears to correlate with infrared flares, albeit with a delay of the sub-mm, suggesting a physical connection between these two regions of the accretion flow, as mentioned in Section~\ref{subsubsec:obs}.
The sub-millimeter fluctuations could  arise from synchrotron emission in the outer regions of the accretion flow, potentially linked to slow-moving turbulence or evolving magnetic fields \citep{2006ApJ...641..302M}. 

\begin{figure}[htb]
    \centering
    \includegraphics[width=0.60\linewidth]{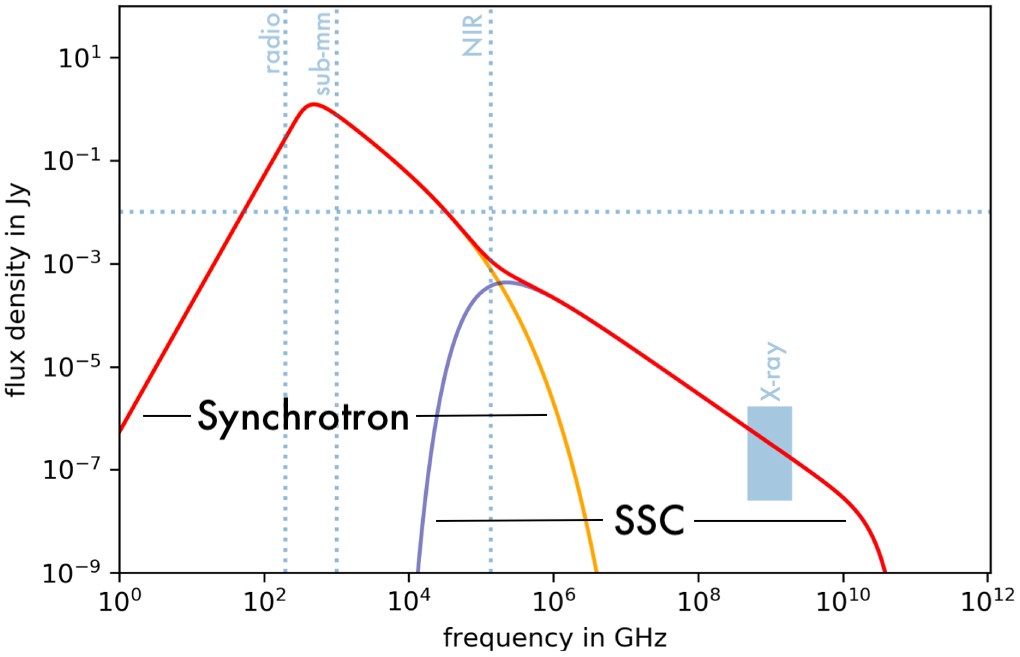}
    \caption{Model radio-to-X-ray spectral energy distribution (SED) for Sgr~ A*. %The vertical dashed lines mark the frequencies 230, 345~GHz, and for the NIR. The blue rectangle marks the 2–8~keV X-ray band and shows typical peak flux densities during X-ray flares. 
    Synchrotron emission (orange) is believed to be responsible for the radio, sub-millimeter, and (largely) infrared photons and their  variability, Synchrotron self-Compton (SSC, blue) emission explains the high-energy X-ray flares, where synchrotron photons from infrared flares are upscattered to X-ray energies. Their sum is shown in red.
    Credit: \cite{2021ApJ...917...73W}.}
    \label{SED}
\end{figure}

Figure~\ref{SED} summarizes the processes invoked to explain the emission at different wavelengths.
While RIAF models provide a good framework for understanding Sgr~A*'s variability, future high-resolution observations, particularly with instruments like the EHT and ALMA, will help refine these models by offering more detailed views of the accretion flow and the physical processes driving the flares.

\subsection{Long term variability}
\label{subsec:longterm}

While the statistical characteristics of Sgr~A*'s short-term variability have not changed much over the past quarter century during which the source has been monitored \citep{ChenZ+19}, there are indications that Sgr~A* was unusually active in 2019 \citep{Weldon+23, 2019ApJ...882L..27D}, possibly due to delayed accretion of tidally stripped gas from the 2014 passage of the extended, dusty G2 object through its orbital periapsis only $\sim$250 AU away from the GBH  \citep{Witzel+14b,Gillessen+19}.  
Other such G objects (a.k.a. dusty stellar objects DSOs, \citealt{2013A&A...551A..18E, 2020Natur.577..337C}) or infalling gas clouds \citep{2023ApJ...944..136C} can also undergo close passages and contribute to the variability on timescales of years.

On timescales of hundreds of years, there are strong indirect indications from X-ray observations that very energetic activity has occurred.  
In many places throughout the Central Molecular Zone, which occupies the Galaxy's central few hundred parsecs, fluorescent X-ray line emission from neutral iron atoms %in many places throughout the Galaxy's few hundred parsec Central Molecular Zone 
has been observed moving away from Sgr~A* at apparent speeds comparable to the speed of light\footnote{Indeed, the apparent speed on the sky plane even exceeds lightspeed in some cases as a result of projection effects.} \citep{Ponti+10,Capelli+12,Clavel+13,Churazov+17,Terrier+18,Khabibullin+22}. 
This has been interpreted in terms of one, or a few, bright flashes of hard X-rays generated by the GBH a few hundred years ago\footnote{An expanding front of hard X-rays can be scattered by the interstellar matter it encounters but, additionally, X-rays can eject the most tightly bound electrons in various atoms, followed by radiative decay of electrons from higher energy states, producing fluorescent X-ray line emission.  The 6.4 keV X-ray line of iron is the most prominent fluorescent line.}, dubbed the "hundred-year events".  
The implied luminosity of these events can be several orders of magnitude greater than even the brightest flares that have been reported since direct X-ray observations of flares from Sgr~A* have been possible (up to several times 10$^{39}$~ergs~s$^{-1}$, \citealt{Porquet+03, Porquet+08, Nowak+12, Haggard+19}).  Furthermore, the durations of the "hundred-year events" are 1 to 10 years -- rather than 0.5 to 2 hours for the brightest directly observed flares -- so the total
energy of the hundred-year event(s) is vastly larger than that of any of the X-ray flares that have so far been directly observed.
The causes of the hundred year events are unknown, but the candidate possibilities include partial tidal disruption of a star, tidal disruption of a planet, or a direct and lasting encounter of Sgr~A* with a dense, compact cloud of gas and dust.

\section{Inflow of Material toward Sgr~A*}
\label{sec:inflow}
The inflow of material toward Sgr~A* occurs on multiple scales and involves complex processes that dictate the dynamics of gas in the surrounding environment. 
Understanding this inflow is crucial for elucidating how black holes accrete matter and influence their host galaxies. 
%The inflow of material toward Sgr~A* occurs on both large and small scales, influenced by the gravitational dynamics of the Galactic Center and the complex behavior of the accretion disk. 
While large-scale inflow provides a reservoir of material, the processes occurring at smaller scales, particularly within the ISCO, dictate the actual amount of material accreting onto the black hole. 
Ongoing observations and theoretical modeling by many groups  are continuously improving our understanding of these inflow mechanisms and their implications for black hole growth and evolution.

\subsection{Large-scale inflow}
On large scales, the inflow of material toward Sgr~A* is primarily governed by the dynamics of the Galactic Center environment. The presence of a dense concentration of stars
%, gas, and dust
within a few hundred parsecs creates a gravitational potential well that influences the motion of surrounding material. Observations have shown that the central few parsecs around Sgr~A* contain a considerable amount of dense molecular gas, which plays a critical role in feeding the black hole.

The CND (see Figure~\ref{fig:GC-cartoon}, left panel), a dense ring of gas and dust surrounding Sgr~A*, has a mass estimated at a few 10$^4$ M$_\odot$ \citep{Requena-Torres+12, Lau+13, Hsieh+21} and is a significant reservoir for potential accretion. %This disk is thought to be structured by gravitational interactions with nearby stars 
This disk is structured by its gravitational interaction with the GBH and with the central concentration of primarily old stars.  
Winds and supernovae from the young nuclear star cluster impact the inner edge of the disk, at a radius of $\sim$1.5 pc, and provoke instabilities that could cause streams of gas to peel off of the disk and fall inward, orbiting closer to the GBH \citep{Blank+16, Solanki+23}. 
This interaction might account for the observed arms of the Mini-spiral, and those gas streams presumably add to the accretion flow onto the GBH (see Figure~\ref{fig:GC-cartoon}, left panel).  
Recent observations using instruments like ALMA and JVLA have provided detailed insights into the distribution and motion of molecular gas in the Galactic Center. 
For example, the presence of high-velocity gas clouds that are being drawn toward Sgr~A* suggests that material from the CND is being accreted onto the black hole, although the process appears to be highly inefficient \citep{2018PASJ...70...85T, Hsieh+19}. 
The inflow rate is currently estimated to be low compared to the mass of Sgr~A*, indicating that while material is available, it is not rapidly falling into the black hole.

%\subsection{Local reservoir of matter}
\subsection{Small-scale accretion}
\label{subsec:smaccretion}
\begin{figure}[htb]
    \centering
    \includegraphics[width=0.65\linewidth]{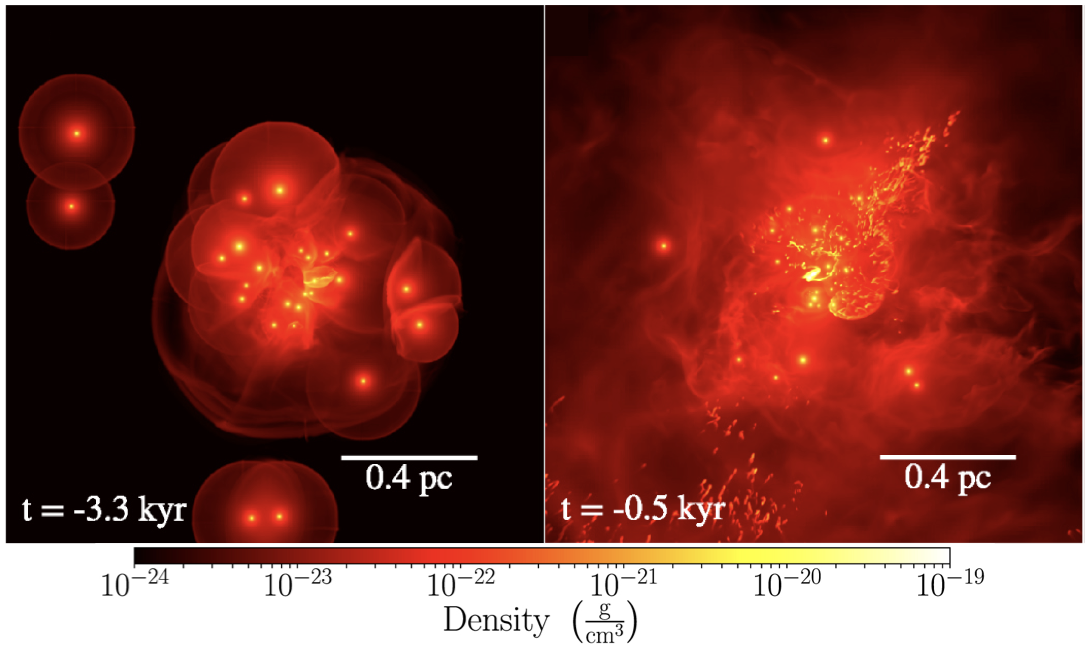}
    \caption{Impact of stellar winds on accretion. The images show two snapshots of a simulation reproducing the interaction of winds emanating from Wolf-Rayet stars in the region and the subsequent chaotic medium they create. Adapted from \cite{2020ApJ...888L...2C}.
    The snapshot on the right shows the formation of an accretion disk at the very center around Sgr~A*; whether such a disk is actually present and detectable at this scale is still under debate \citep{2019Natur.570...83M, 2021ApJ...910..143C}.}
    \label{fig:WR}
\end{figure}

At smaller scales, the dynamics of material inflow become significantly more intricate. 
At present, winds emanating from young Wolf-Rayet stars within 0.05 to 0.5 parsecs from Sgr~A* \citep{Paumard+06} represent the main source of accreting material (see Figure~\ref{fig:WR}, \citealt{Ressler+18, 2020MNRAS.492.3272R,2020ApJ...888L...2C}).
These potent stellar winds constantly replenish the central parsec with hot ($\sim10^7~$K) and diffuse plasma as shown by Chandra observations \citep{Baganoff+03, 2013Sci...341..981W}\footnote{the central parsec, while filled with hot plasma and strong ultraviolet radiation, is not devoid of material: diffuse dust and gas are present, arranged in a very clumpy medium \citep{2016A&A...594A.113C, 2019A&A...621A..65C}.}. 
The total mass-loss rate of the Wolf-Rayet stars is roughly 10$^{-3}$ solar masses per year \citep{2007A&A...468..233M} but 
the accretion rate through a radius of hundreds of Schwarzschild radii (10$^{-5}$~pc) is orders of magnitude smaller (10$^{-9}$--10$^{-7}$ solar masses per year, as measured by polarized emission at submillimeter wavelengths \citep{Bower+03, 2006ApJ...640..308M}.
Although the present-day accretion rate is extremely low, it has apparently been higher at various times in the past (see Section~\ref{sec:outflow}). 

Within the ISCO, the dynamical behavior of  the gas is dominated by extreme general relativistic effects, as well as by relatively strong magnetic fields. %The region within the ISCO is characterized by extreme gravitational effects, where the behavior of gas and dust is dominated by relativistic physics and relatively strong magnetic fields. 
Here, the inflow is largely influenced by the properties of the accretion disk.
The accretion flow near Sgr~A* is often described by the RIAF model (see Section~\ref{models}), which posits that much of the material entering the black hole does so without radiating significant energy. 
In RIAF models, the inflow is governed by viscous processes and turbulence within the accretion disk, leading to a low radiative efficiency \citep{1995ApJ...452..710N, 2014ARA&A..52..529Y}. Advection-Dominated Accretion Flow (ADAF) model is one form of RIAF model in which most of the energy -- gained by inspiralling material as it converts
gravitational potential energy to kinetic energy -- is carried inward across the event horizon rather than being radiated away. 
%which has the characteristics described in Section~\ref{subsec:outflowmodels}.
%As material spirals inward, it becomes increasingly compressed and heated, emitting radiation primarily in the infrared and sub-millimeter bands due to synchrotron processes (see also Section~\ref{models}).

Recent observations have highlighted the presence of hot spots in the accretion flow, which are regions of concentrated material that orbit close to the ISCO. These hot spots can lead to the enhanced emission observed as flares \citep{GravCollab18b, Ripperda+22}.
The behavior of these hot spots provides critical insights into the accretion processes at play, as they reflect the interplay between gravitational forces and magnetic fields in the inner regions of the accretion disk.
Furthermore, theoretical models suggest that fluctuations in the accretion flow can lead to instabilities, resulting in rapid inflow of material during certain episodes. These episodes could be driven by tidal interactions or magnetic reconnection events, causing sudden increases in the inflow rate. 
As accreting material approaches the black hole, it transfers some of its angular momentum to the outflow, allowing that material to spiral inward until %lAs material approaches the black hole, it loses angular momentum, causing it to spiral inward until 
it reaches the ISCO, where it is no longer in a stable orbit and will inevitably fall into the black hole.

%\subsection{Accretion models}

\section{Outflow}
\label{sec:outflow}
In addition to accreting material, the GBH also produces outflows. %that impact its surrounding environment. Outflows are an essential aspect of black hole physics as they carry energy and angular momentum away from the accretion flow, significantly affecting the Galactic Center's dynamics and regulating the growth of the black hole itself (for example, by impeding accretion when the outflow becomes strong enough). 
The outflow from Sgr~A* occurs across multiple scales and appears to be relatively weak compared to the powerful jets seen in more active galaxies. 
Nonetheless, 
%%% Moved from "impact of outflow"
the outflows from Sgr~A* can significantly impact the surrounding environment. %, even if they are not as powerful as those observed in more active systems. 
It is a curious fact that the vast majority of the matter that is drawn toward the black hole and participates in the accretion flow eventually gets expelled, rather than actually going through the event horizon and adding to the black hole mass. 
There are several contributing factors for this, including the conservation of angular momentum, strong heating of the accreting matter, the dynamical role of magnetic fields, and violent instabilities in the rapidly evolving accretion disk. 

By carrying away angular momentum and energy, these outflows help regulate the accretion process and limit the amount of material that reaches the black hole. This feedback mechanism is believed to be critical in maintaining Sgr~A*'s low luminosity and low accretion rate.
Moreover, the outflows interact with the CND and the surrounding interstellar medium, potentially influencing star formation in the region. The energy carried by these outflows can heat nearby gas, preventing it from cooling and collapsing to form stars. Alternatively, they can compress gas clouds, triggering star formation in certain conditions. Observations of shocked gas and complex velocity structures in the Galactic Center suggest that the interplay between outflows from Sgr~A*, along with the strong winds from the massive stars in the YNC, and the surrounding medium plays an active role in shaping the dynamics of the region \citep{1999AdSpR..23..959M, 2013Sci...341..981W}.
%%%%

%Nonetheless, these outflows play a crucial role in regulating the accretion process, expelling angular momentum, and interacting with the surrounding medium. Understanding the mechanisms driving these outflows, as well as their impact on the Galactic Center environment, is key to forming a complete picture of Sgr~A*'s behavior and its influence on the Milky Way.

\subsection{Observational Evidence for Outflows}
\label{subsec:outflow}
\begin{figure}[htb]
    \centering
    \includegraphics[width=\textwidth]{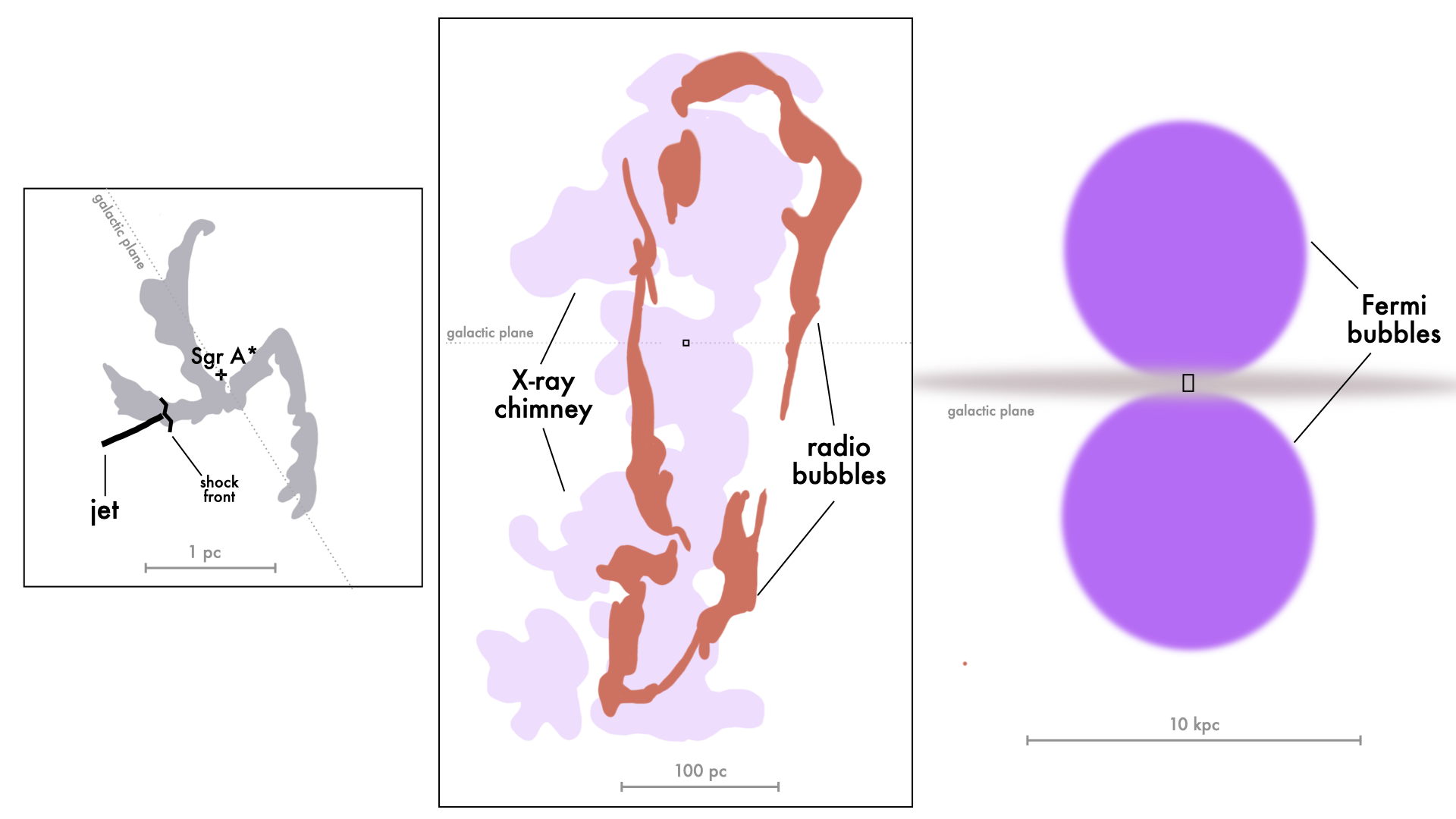}
    \caption{Cartoon illustrating the main observational evidence of outflows from the Galactic Center that can be found at different scales. The little squares on the mid and right panels illustrate the scale of the panel to the left of each, respectively.}
    \label{fig:outflows}
\end{figure}

Direct evidence for outflows from Sgr~A* comes from multiple observations across different wavelengths (see Figure~\ref{fig:outflows}). 
%Radio and X-ray observations, in particular, have revealed structures consistent with the presence of outflows or jets. 

While Sgr~A* does not exhibit the powerful relativistic jets often seen in more active galactic nuclei on kilo-parsec scales, there is evidence that the GBH has launched an X-ray emitting jet detectable on scales of a few parsecs \citep{Li_2013}. 
It manifests itself as a linear feature aligned with the GBH (see Figure~\ref{fig:SgrA*_radio}, right panel and Figure~\ref{fig:outflows}, left panel), and has a spectrum that steepens with increasing distance from Sgr~A* \citep{ZhuZhenlin+19}, as would be  expected because the emitting particles progressively lose energy as they emit and move outward. 

There are also indications of weaker, collimated outflows from the GBH.  
Radio observations using the JVLA have identified elongated features extending away from Sgr~A* on scales of 10-20 pc, suggesting the presence of a mildly collimated outflow \citep{ZMG16}. These features, sometimes referred to as "lobes", are also manifested in X-ray emission \citealt{Morris+03, Ponti+15}.  
These radio/X-ray features are thought to be the result of synchrotron emission from relativistic electrons that have been accelerated by magnetic processes in the inner accretion flow. 
The lobes have possibly resulted from past sudden increases in the accretion rate onto the GBH, but they could also have been produced, or enhanced, by supernova explosions of one or more massive stars in the Young Nuclear Cluster. 

On scales of several hundred parsecs, X-ray and radio astronomers have recently discovered quasi-cylindrical "chimneys" rising away from the central $\sim$100~pc around Sgr~A*, on both sides of the Galactic plane \citep{Ponti+19,Ponti+21,Heywood+19}. The X-ray morphology strongly suggests that a very hot plasma is rising through the chimneys, forming what might be an "exhaust vent" from the energetic activities in the Galactic center. The radio emission forms "bubbles" that apparently surround the bulk of the X-ray emitting plasma (see Figure~\ref{fig:outflows}, middle panel).  The source of the hot plasma could be some combination of outflows from sizable accretion events onto the GBH over the past 10$^5$ years, including stellar tidal disruption events, and relatively frequent supernovae taking place in the Galaxy's Central Molecular Zone \citep{ZhangMF+21}.  One deep X-ray image of the southern chimney shows shock fronts that indicate that the outward flow velocity of the plasma in the central conduit of the chimney might be rather high \citep{Mackey+24}. The placement and size of the chimneys raises the possibility that they are  contributing to the inflation of the Fermi Bubbles, perhaps by transporting cosmic rays out from where they are formed near or around Sgr~A*.
%These outflows may carry a significant amount of mass and energy away from the black hole, contributing to the feedback processes that regulate its activity.

%This hot gas is ejected at sub-relativistic velocities and may form part of a larger-scale wind that influences the dynamics of the surrounding medium.
%Additionally, observations in the infrared have indicated the presence of ionized gas flowing away from the black hole, possibly driven by the interaction of hot spots and magnetic fields in the accretion disk. The GRAVITY instrument on the VLTI has provided evidence for small-scale outflows that might be associated with magnetic reconnection events or turbulence within the accretion flow \citep{2018A&A...618L..10G}.

On a much larger scale (10~kpc), the Fermi Bubbles represent a striking outflow feature carrying very high energy relativistic particles.
They are manifested as giant, well-defined gamma-ray-emitting bubbles found by \citet{Su+10} using data from the Fermi Gamma-Ray Space Observatory (see Figure~\ref{fig:outflows}, right panel).  
The Fermi Bubbles are symmetrically placed above and below the Galactic plane, extending about 8~kiloparsecs into the Galactic halo, with their shapes converging on the Galactic Center. 
They could have been created by a strong, collimated wind from the GBH during a high-accretion AGN stage \citep[see Section~\ref{sec:starformation},][]{Zubovas+11, Zubovas+12b, Yang+22}.
However, there are alternative interpretations of how the Fermi Bubbles could have been produced on much longer time scales \citep{Crocker+15, Yang+18}, so more work is needed to definitively distinguish between these hypotheses.

\subsection{Theoretical Models of Outflows}
\label{subsec:outflowmodels}
Matter in the accretion flow invariably has angular momentum, and therefore cannot simply fall directly onto the black hole.  Instead, the infalling accretion flow reaches a centrifugal barrier, where instead of falling further, it goes into orbit, and can only spiral further inward by losing angular momentum.  The accreting matter "circularizes" at that point, forming an accretion disk, the detailed structure of which depends on whether the accretion flow is ordered (e.g., spherically or axially symmetric), time-variable, or chaotic. 
Continuing rapid infall of material onto the slowly inspiralling accretion disk leads to strong shocks \citep{MeliaFalcke01} that heat the gas to temperatures at which the gas is hot enough to emit thermal X-rays. The circularization and the resulting shocks occur near the Bondi capture radius, which in the case of the GBH is at an angular radius of 1--1.5~arcseconds\footnote{large enough for the Chandra X-ray Observatory to have resolved the resulting, relatively constant  "quiescent" X-ray emission \citep{Baganoff+03,2013Sci...341..981W}.}.  
Because of viscous evolution of the accretion disk, mediated by both turbulence and the magnetic field, angular momentum is transferred outward, and the outer layers of the accretion disk must therefore expel matter. But throughout the accretion disk, as the material approaches the GBH, the magnetic field strength and the gas temperature both rise to very high values.  
Through complex processes still to be elucidated in detail, the coupling of the magnetic field and the rotation of the disk conspire to drive a wind off of the disk, thereby removing most of its mass and angular momentum.  
That allows some of the disk material to finally reach the innermost portions of the disk, close to the ISCO, where the gas has become so heated that the particles within it move relativistically,  emitting non-thermal radiation.  

Theoretical models for the outflow from Sgr~A* generally fall within the framework of magnetohydrodynamic processes and RIAF (see also Section~\ref{models}). 
In these models, outflows are driven by magnetic fields threading the accretion disk, which can accelerate and channel material away from the black hole. 
The interaction between the black hole's spin and the magnetic fields can also drive outflows, as described by the Blandford-Znajek mechanism \citep{1977MNRAS.179..433B}.
In the case of Sgr~A*, which is relatively inactive compared to more luminous active galactic nuclei, outflows are believed to be weaker and less collimated. This is partly because the accretion flow around Sgr~A* is apparently advection-dominated, as described in Section~\ref{subsec:smaccretion}. Consequently, the resulting outflows are less energetic and more diffuse, often appearing as weak winds rather than focused jets \citep{1995ApJ...452..710N}.
%Another important theoretical model is the disk-wind model, which suggests that the accretion disk itself can launch outflows due to centrifugal forces acting on magnetized plasma. In such scenarios, magnetic field lines anchored in the disk act as channels along which gas is expelled. The outflows produced by this mechanism are slower and less collimated compared to jets %produced by the black hole's rotation 
%but still play an essential role in regulating the mass inflow \citep{2014ARA&A..52..529Y}.

%%% moved from intro

%The primary consideration is that the 
%%%%%

\section{Summary and Perspectives}
The detailed and extensive observations of the GBH conducted over the past decades have provided significant insights into the behavior of black holes and their interactions with the surrounding environment.
\begin{itemize}
    \item \textbf{Testing General Relativity:} The Galactic Center provides one of the best laboratories to test the predictions of general relativity in the strong gravitational field of a supermassive black hole. Observations of stellar orbits, particularly S0-2/S2, have confirmed effects like gravitational redshift and Schwarzschild precession, consistent with general relativistic predictions. Future observations may also reveal more subtle effects, such as frame dragging, which would provide additional tests for Einstein's theory.

    \item \textbf{Variability and Accretion:} Sgr~A* shows variability across all wavelengths, from radio to X-rays, with flare events providing a window into the dynamic processes occurring in the accretion flow. Observations indicate that the variability follows a power-law distribution and is likely driven by turbulent, transient processes such as magnetic reconnection or hot spot formation. Processes like RIAF and SSC emission %(see Section~\ref{models})
    help explain the observed variability across different wavelengths.

    \item \textbf{Inflow and Outflow Processes:} The accretion of material onto Sgr~A* involves a complex interplay of large-scale and small-scale processes. Material flows inward from the circumnuclear disk at large scales, whereas near the black hole, the inflow is shaped by the properties of the accretion disk and the conditions near the innermost stable circular orbit. Simultaneously, outflows, though weaker compared to the powerful jets of more active black holes, are nonetheless an important component that helps regulate accretion and impacts the Galactic Center environment by redistributing energy and momentum.

\end{itemize}

\subsection{Future instrumentation}
The next decade or two promises to be transformative for our understanding of Sgr~A*, with advancements in instrumentation, observational techniques, and theoretical modeling.
Upcoming observatories like the Extremely Large Telescope (ELT) and Thirty Meter Telescope (TMT) will provide unprecedented spatial resolution and sensitivity, enabling the detection and monitoring of stars even closer to Sgr~A*. These observations will help refine our understanding of the black hole's mass, the accretion process, and the dynamics of the innermost regions of the Galactic Center.
The James Webb Space Telescope (JWST) will add significantly to our understanding by providing high-sensitivity infrared observations, which will help penetrate the dense dust in the Galactic Center and observe fainter stars that are too dim to be seen with current ground-based telescopes.
%Instruments like the GRAVITY interferometer at the VLTI will continue to play a crucial role in tracking the positions of stars near Sgr~A* with sub-milliarcsecond precision, contributing to more accurate measurements of stellar orbits and further testing general relativity.
Improvements in sensitivity and time sampling will make it possible to detect more stars closer to the black hole, which may show frame dragging effects caused by the black hole's spin. % (also known as the Lense-Thirring effect). 
Detecting and measuring this effect would provide critical insight into the spin properties of Sgr~A*, further testing general relativity in extreme conditions.
Future millimeter and submillimeter observations, particularly with ALMA and the EHT, will improve our understanding of the small-scale structure of the accretion flow and the formation of jets or outflows. These observations will help constrain models such as RIAFs and provide insights into the physics behind the variability observed in different wavelengths.
Observations will also focus on the link between the inflow and outflow mechanisms, examining how the black hole's accretion rate affects the properties and strength of the outflows. This will enhance our understanding of how feedback works on both small and galactic scales.
Observing the interaction between Sgr~A*'s outflows and the surrounding circum-nuclear disk will shed light on how the black hole affects the local environment, including star formation. With new instruments, we will be able to better understand whether these outflows trigger or suppress star formation in the Galactic Center, contributing to our understanding of the co-evolution of black holes and galaxies.
Multi-wavelength campaigns, combining data from radio, infrared, and X-ray observatories, will continue to provide complementary insights into the variability and flare activity of Sgr~A*. The development of more sophisticated theoretical models, supported by the increasing observational data, will help clarify the processes behind the variability, the origin of flares, and the coupling between different regions of the accretion flow.

\subsection{Conclusions}
In the past half-century, Sgr~A* has evolved from being a mysterious radio source to becoming a well-characterized supermassive black hole that offers a wealth of information about black hole physics, accretion, and relativistic effects. 
With the next generation of observatories, both on the ground and in space, we are entering an era where we can probe the dynamics of the Galactic Center in greater detail than ever before. 
These advancements promise to not only answer existing questions but also reveal new, unexpected phenomena, helping us to understand how SMBHs were formed, how they grew over time, and the impact that they have had on their environments. %the true nature of supermassive black holes and their role in the cosmos.
Since almost all sufficiently large galaxies appear to harbor a central massive black hole, the GBH is a proxy for elucidating black-hole-related physics throughout the universe. 

%\section{Level 1 heading}\label{chap1:sec1}
%\subsection{Level 2 heading}\label{chap1:subsec1}
%\subsubsection{Level 3 heading}\label{chap1:subsubsec1}
%\paragraph{Level 4 heading}
%\subparagraph{Level 5 heading}%
%\subsubparagraph{Level 6 heading}%
%\subsubsubparagraph{Level 7 heading}%

\begin{ack}[Acknowledgments]

We are grateful to the UCLA Galactic Center Group for supporting our research on the topics discussed here, as well as on other Galactic Center science, creating a dynamic environment in which research on this topic is active and ongoing.
\end{ack}

\seealso{For additional details on Sgr~A* see also reviews by \cite{2023arXiv230202431M, 2010RvMP...82.3121G}, and for a comprehensive review of the Galactic Center region see reviews by \cite{2021NewAR..9301630B, 1996ARA&A..34..645M}.}

\bibliographystyle{Harvard}
\bibliography{reference}

\end{document}